\documentstyle [12pt]{article}
\newcommand{\be}{\begin{equation}}
\newcommand{\ee}{\end{equation}}

\begin{document}

\title{Intermittency and eddy-viscosities in dynamical
models of turbulence}

\author{ R. Benzi$^{1}$, L. Biferale$^{2,3}$, S. Succi$^4$ and
F. Toschi$^{3,5,6}$\\
\it $^1$AIPA, Via Po 14, 00100 Roma, Italy.\\
\it $^2$Dipartimento di Fisica, Universit\`a "Tor Vergata"\\
\it Via della Ricerca Scientifica 1, I-00133 Roma, Italy.\\
\it $^3$INFM, Unit\`a di Tor Vergata.\\
\it $^4$Istituto Applicazioni Calcolo,CNR\\
\it V.le Policlinico 137, 00161, Roma, Italy\\
\it $^5$Dipartimento di Fisica, Universit\`a di Pisa\\
\it P.zza Torricelli 2, I-56126 Pisa, Italy\\
\it $^6$INFN, Sezione di Pisa\\
}
\maketitle

\begin{abstract}
The dependence of intermittent inertial properties 
on ultraviolet eddy viscosity closures
is examined within the framework of shell-models of turbulent flows.
Inertial intermittent exponents turn out to be
fairly independent on the way energy is dissipated at small scales. 

\end{abstract}
\section{Introduction}
One of the most challenging open problems in three dimensional fully
developed turbulence is the assessment of the statistical properties 
of the energy transfer mechanism.
In stationary turbulent flows, a net flux of energy establishes in the 
inertial range, i.e. from forced scales, $L$,  down to the
dissipative scale, $r_d$. 
Energy is transferred through a statistically scale-invariant process,
characterized by a strongly non-gaussian (intermittent) activity.\\
Intermittency is usually described by looking at the statistical
properties of longitudinal velocity differences, $\delta_r v(x) =v(x) -v(x+r)$ (vector notation is relaxed for simplicity). 
In particular, the last twenty years \cite{frisch} have witnessed a substantial
focus of experimental and theoretical activity on structure functions:
$
S_p(r) = <(\delta_r v(x))^p> $. A wide consensus exists on the fact
that structure functions show a scaling behavior in the limit of
very high Reynolds numbers, i.e. in presence of  a large
separation between the integral and dissipative scales, 
$L/r_d \rightarrow \infty$:
\be
S_p(r) \sim \left({r \over L} \right)^{\zeta(p)}.
\label{structure}
\ee
The velocity fluctuations are anomalous in the sense that the
$\zeta(p)$ exponents do not follow the celebrated dimensional
Kolmogorov's prediction $\zeta(p)=p/3$.
In fact, $\zeta(p)$ is observed to be a nonlinear
function of its argument $p$, which is interpreted as 
the most important signature of the intermittent transfer of fluctuations 
from large to small scales.\\
As it is known, the dissipative structure of the Navier-Stokes
equations (NS) is not dictated by compelling constraints on the 
inertial terms. This raises the question 
on whether or not the statistical properties of fully developed 
three dimensional turbulent flows exhibit a 
strong dependency on the energy dissipation mechanism.

Kolmogorov theory suggests a strong
universality assumption: strong independency
of pure inertial range quantities on any dissipative
mechanism. 
The theoretical implication of such an assumption are obvious. 
For instance, some of the most recent analytical
attacks to the intermittency of structure functions 
assume that the phenomenon is fully captured
by looking only at the nonlinear terms in the NS eqs, at least in the
limit
of large Reynolds number \cite{procaccia}. However, because of
intermittency, one can question the conceptual framework of the
Kolmogorov theory and consequently the strong universality assumption.\\
Moreover, numerical investigations of turbulent flows are necessarily
restricted to low (moderate) Reynolds numbers.  Therefore, it is of
primary importance to develop some controllable procedure which minimizes
viscous effects (whenever possible). \\ 
In the past, hyperviscosity (high powers of the Laplacian) has often
been employed in order to extend the inertial range as much as possible. 
Contradictory claims have been reported on the influence 
of the energy dissipation mechanism
on the inertial range dynamics \cite{orszag2d,chen,she,kadanoff,eyink}.

An important tool, heavily in use to perform reliable high-Reynolds simulation, 
is based on the concept of eddy viscosity \cite{frisch,refeddy1}.\\
In this paper we investigate robustness of the intermittent
inertial properties in the context of simple dynamical eddy viscosity models. 
In particular, we  present a detailed numerical investigation
of such an issue in a class of dynamical models of turbulence (shell models)
both in the case where the dynamics is resolved in real and Fourier space 
(tree model)\cite{bbtt,erik}
and in the case where only the Fourier space is taken into
account (standard chain-models \cite{GOY,JPV,KLWB,BK,bbkt}).\\
In either cases, we found strong independence of the inertial range
statistics from the ultraviolet dynamical closure, indicating that 
most (eventually not all) eddy viscosity models do not destroy 
the quantitative and qualitative features of the inertial range dynamics.\\
The paper is organized as follows. In section 2 we introduce the
main ideas behind eddy viscosity models.
In section 3 we introduce the dynamical models we have used in order to test the
dependency of intermittency on eddy viscosities.
In section 4 we discuss the numerical results. Conclusions follow in section 5.

\section{Eddy viscosity models}
The idea of eddy viscosity was introduced over a century ago by Boussinesq and later 
developed further by G. Taylor and L. Prandtl \cite{FRIORS} and it builds upon a direct
 analogy with the kinetic theory of gas. According to this analogy, the effect of short 
'microscopic' scales on large 'macroscopic' scales can be likened to a sort of diffusion
process characterized by an effective viscosity much larger than the molecular one.
Strictly speaking, this is justified only when a sharp separation
between 
fast and slow
modes exists, but it turns out that the analogy proves useful in practice 
also in situations where, in principle, such an assumption would not hold.

By mere dimensional arguments, the effective eddy viscosity 
at scale $r$ reads as follows 
\be
\label{EDDY}
\nu_{\tiny E}(r) \sim r \cdot \delta v (r)
\ee
where $\delta v (r)$ is the velocity fluctuation across a distance $r$
 (vector indices are
relaxed for simplicity).\\
Equation   (\ref{EDDY}) can be also deduced by using 
 the refined Kolmogorov hypothesis (RKH) as follows.
According to Kolmogorov \cite{frisch}, a simple way to
take into account
the intermittent fluctuations in the inertial range is to define a
coarse grained energy 
dissipation $\epsilon_r(x)$:
\be
\epsilon_r(x)= {1 \over {r^3}} \int_{\Lambda_r(x)} \epsilon(y) \; d^3y
\ee
where $\Lambda(x)$ denotes a volumlet of fluid centered in $x$.

In terms of $\epsilon_r$ one can
generalize the Kolmogorov "4/5" equation by assuming that $ (\delta_r
v(x))^3 \sim r \; \epsilon_r $.\\
Now, let us define $\Delta$ the scale at which we want
to compute the
eddy viscosity. At such a scale, one expects $\epsilon_{\Delta} = 
\nu_{\tiny E}( \Delta)  \left({{\delta v (\Delta)} \over {\Delta}}\right)^2$.
By combining these two expressions (\ref{EDDY}) is readily obtained.\\
The eddy viscosity is much larger than the molecular one, which reflects
the enhanced mass and momentum transport observed in turbulent flows.\\
As it is well known, for most turbulent flows of practical interest,
the dissipative scale $\eta$ is far too short to be resolved by 
any foreseeable computer. 
In fact $\eta$ scales like $L \cdot Re^{-3/4}$, $L$ being the outer scale 
of the flow, and consequently the scale separation $L/\eta$ can easily
span $3-6$ orders of magnitude in practical applications. \\
Given this state of affairs, subgrid models and large-eddy-simulations 
(LES), are mandatory. 
Generally speaking, the common aim of these models is to incorporate
the effects of unresolved scales ($r<\Delta$, $\Delta$ being a typical
mesh size) on the resolved ones, $r>\Delta$.

One of the simplest and most popular sub-grid-model is due 
to Smagorinski \cite{smago}, which can be derived by (\ref{EDDY}).\\
The idea is to replace $r$ with the mesh size $\Delta$ in 
the eq. (\ref{EDDY}) and  
subsequently replace $\delta v (\Delta) \sim S \Delta$, where
(we dispense from tensor indices for the sake of the argument) 
$S$ is the strain tensor $S \sim \delta v/r$ evaluated at $r=\Delta$.\\
The result is
\be
\label {SMAGO1}
\nu_{\tiny{SGS}} \sim \Delta^2 S 
\ee
This expression is less transparent than it looks. In fact, it is based
on the assumption that the velocity field at the scale $\Delta$ is
smooth
enough to allow the definition of the space derivative $S$. 

This flies in the face of the fact that, if $\Delta$ belongs to the
inertial
range (as it should for the whole LES procedure to make sense), 
the velocity field is {\it known} not to be differentiable
since $\delta v$ scales like $r^{1/3}$.
On account of this, one expects $\delta v(\Delta)/\Delta$ be much larger
than the corresponding ratio evaluated at $r=\eta$ (the only scale
where
this operation is conceptually allowed).
This 'inconsistency' is usually acknowledged by prefactoring the
right hand side of the equation  with an empirical
coefficient $C_S$ smaller than one, typically $C_S \sim 0.12$.\\
Putting all together, and restoring tensorial indices, the full
Smagorinski eddy-viscosity reads as
\be
\nu_{\tiny{SMG}}(x) = C_S \; \Delta^2 \left| S \right| S_{ij}.
\ee
where $S_{ij} = {1 \over 2} \left( \partial_i u_j + \partial_j
u_i\right)$ 
is the large-scale stress tensor and  
$\left| S \right| = \left( 2 S_{ij} S_{ij} \right)^{1/2}$.\\
The Smagorinski model is widely used in practical engineering
applications
in spite of its several weaknesses.
Among these, worth mentioning are i) the overdamping of resolved scales,
and
ii) the, at least partial, assumption of isotropy of the turbulent
flows.
The former flaw may seriously behinder the development of genuine
instabilities
\cite{PIO}, while the latter casts doubts on the applicability of the
model in
the vicinity of walls and solid boundaries where the dynamics of
turbulence is
dominated by directional effects.

Another recent development in the area of Smagorinski  models is 
the so-called
structure-function eddy-viscosity by Lesieur \cite{LESIEUR1}
and coworkers.
The idea is to account for intermittency by estimating $\delta v(r)$ 
with the square root of the second order local structure function
\be
\label{LESIEUR}
\delta v (r) \sim  S_2(r,x)^{1/2} \equiv \left\langle \delta v(r,x)^2 \right\rangle^{1/2}
\ee
where the local average is computed with the local energy spectrum 
$E(k,x,t)$ according
to the Batchelor  relation
\be
\label{BATCH}
S_2(r,x) = \int E(k,x) \frac{\sin(kr)}{r} dk.
\ee
The relation (\ref{LESIEUR}) implies a certain degree of
arbitrariness. 
Why not choosing $\delta v \sim S_3^{1/3}$ or more generically
$S_p^{1/p}$ with $p$ any integer? 
In the absence of intermittency all $p$'s are equivalent, but 
when intermittency is on, every value of $p$ would provide a 
different, yet equally acceptable, answer.
At this stage, the specific choice of $p$ becomes a matter
of taste, or, better said, of how much emphasis is to be placed 
on the most-singular structures (those sampled by highest $p$'s).
The correct recipe is probably a weighted average of all possible $p$'s,
the weighting factor (most likely a space-time dependent function) 
being basically unknown.\\
Another scenario is to assume that intermittency
ignores the details of the dissipation mechanisms, in which case the
idea of including intermittency effects on LES models dissolves on its own.
\\
The discussion of the sophisticated developments of LES modeling is
beyond the scope of this paper, here we shall focus exclusively
on the specific question of the interrelation between dissipation and 
intermittency.
Tackling this question within the true tree-dimensional Navier-Stokes 
context is a very daunting task, in view of the enormous amount of data
to be produced and carefully analyzed.\\
It makes therefore sense to attack the problem within the context of
simplified dynamical models sharing as much physics as possible with
Navier-Stokes equations while giving away most of its computational complexity.
  
\section{Dynamical models of turbulence}
In the recent years, an interesting  vehicle for this kind of
investigations
has emerged in the form of the so-called ``shell models'' \cite{GOY}-\cite{bbkt}.\\
Shell models work on the principle of collapsing the whole set of
degrees
 of freedom lying in a finite shell $k_n<k<k_{n+1}$, with $k_n = 2^n k_0$,
 into a handful (one, two)
  of representative modes.\\
The dynamics of such a low-dimensional representation is subsequently
arranged
 in such a way as to preserve the non-linear structure of the NS equations; 
 of course
  all genuinely three-dimensional effects are lost in the process.\\
The most popular shell model is the Gledzer-Ohkitani-Yamada (GOY) model
where 
only one (complex) mode per shell is used. 
Recently, a new class of model has been introduced in which, by allowing
two
 complex modes per shell, a second invariant with a 
  close connection to NS helicity can be defined.\\
The statistical properties of such a helical shell model have been
recently explored in depth \cite{bbkt,bbt}, 
major finding being that it possesses a rich physics and it
exhibits a striking similarity 
(in a statistical sense) with NS  intermittency. 
Shell models do nonetheless miss all spatial effects, since they can be 
regarded as zero-dimensional field models based on space-filling coherent
planar waves. \\
The dynamics of our helical  shell model is governed by the following 
evolution equation:
\be
\label{hel}
{\dot u}_n^{\pm} = i k_n \left( a u_{n+1}^{\pm} u_{n+2}^{\mp}
+ b u_{n-1}^{\pm} u_{n+1}^{\mp}
+ c u_{n-1}^{\pm} u_{n-2}^{\mp}
 \right) - D^{\pm}_n  u_n^{\pm} + \delta_{n,n_0}f^{\pm}
\ee
where $u_n^{\pm}$ represent the positive/negative helicity 
carriers respectively and $f^{\pm}$ is a large scale forcing.
In the previous equations the term  $D^{\pm}_n$ is a function which reproduces
the effects of viscous damping at scale $n$. In the usual case where
only molecular viscosity, $\nu$, is acting we have 
$$ D^{\pm}_n =\nu k_n^2.$$
 Upon choosing 
$a=1, b= - {5 \over 12}, c= - {1 \over 24}$, the above
 equations are readily shown to 
conserve the following (energy, helicity) invariants
\be 
E = \sum_{n=0}^N \left(|u_n^+|^2+|u_n^-|^2 \right) ;\;\;\;
H = \sum_{n=0}^N k_n \, \left(|u_n^+|^2-|u_n^-|^2 \right).
\ee
Real turbulence consists of localized eddies of all sizes that interact,
merge and subdivide locally: the physical picture is that of a 
large eddy  which decays into smaller eddies. 
The number of degrees of freedom in such a field problem in
$d$ dimensions grows with the wave number as $N(k) \sim k^d$
($d=0$ in  shell models).
The first step in reproducing 
this kind of hierarchical structure is to transform  a  {\it
chain}-model 
 into a {\it tree}-model with $d=1$ \cite{bbtt}.
This is achieved by letting the number of degrees of freedom
grow with the shell index $n$ as $2^n$.

As in the original shell models, this tree model
must be in some sense reminiscent of the NS equations.
It can be regarded as 
describing the evolution of the coefficients of an orthonormal wavelets 
expansion of a one-dimensional projection of the 
velocity field $v(x,t)$:
\begin{equation}
v(x,t)=\sum_{n,j} \hat{v}_{n,j}(t) \psi_{n,j}(x).
\label{eq:wave}
\end{equation}
Here $\psi_{n,j}(x)$ are a complete orthonormal set of wavelets 
generated from
an analyzing wavelet $\psi_{0,0}(x)$ by discrete translations and 
dilatations:
\begin{equation}
 \psi_{n,j}(x)= 2^{n/2} \psi_{0,0}(2^n x-j).
\label{eq:wave1}
\end{equation}

Each dynamical variable $\hat{v}_{n,j}$ describe fluctuations 
in a box of length $l_n=2^{-n}$, centered in the interval  ranging 
from $(j-1)l_n$ to $j l_n$.
At each scale $n$ there are $2^{n-1}$ boxes, covering a total length 
$ \Lambda_T=2^{n-1} l_n=1/2 $ (see Fig. 1).

For the sake of convenience we define the tree model in terms of {\it
density} variables,
$u_{n,j}$, which would cor\-re\-spond to  
$\hat{u}_{n,j}= 2^{n/2} \hat{v}_{n,j}$ in a wavelets expansion.\\
In this notation,  
$| u_{n,j} |^2$ 
represents the energy density in a flow structure of length
$l_n=2^{-n}$ and spatially labeled by the index  $j$.\\
In this tree structure, each variable $u_{n,j}$  
 continues to interact with the nearest and next nearest
levels, as in equation (\ref{hel}); however, a lot
 of possibilities are now opened by the presence of many 
 horizontal degrees of 
freedom localized on each shell.\\
 The simplest choice is depicted in Fig. 2, where a portion of 
 the tree structure is shown and the evolving in time variable, 
  $u_{n ,j}$,
 is represented by a black ball.
 In the figure, solid lines connect interacting balls (variables).\\
The dynamical tree equations are as follows:
\begin{equation}\begin{array}{ll}
\dot{u}^+_{n,j}=  - D^+_{n} u^+_{n,j} +\delta_{n,n_0} F^+ +\nonumber\\
+ i k_n  \left\{{a \over 4}\; \left[ u^{+}_{n+1 ,2j-1}\, \left(u^{-}_{n+2 ,4j-3} + u^{-}_{n+2 ,4j-2}\right)+ u^{+}_{n+1 ,2j}\, \left(u^{-}_{n+2 ,4j-1} + u^{-}_{n+2 ,4j}\right) \right] \right. +\nonumber\\
\left. + {b \over 2} \left[u^{+}_{n-1,\bar{\it \j}} \, \left(u^{-}_{n+1,2j-1}+u^{-}_{n+1,2j}\right)\right] + c \left[u^{-}_{n-2,\bar{\bar{\it \j}}} \, u^{-}_{n-1,\bar{\it \j}} \right] \right\}^* 
\label{TREE}
\end{array}
\end{equation}
where, in the indexes, $\bar{\bar{\it \j}}$ is the integer 
part of $\left(\frac{j+3}{4}\right)$ and 
$\bar{\it \j}$ is the integer part of $\left(\frac{j+1}{2}\right)$.\\
Again, in the standard case with only molecular viscosity we have 
$D^{\pm}_n=\nu k_n^2$. The interaction terms with coefficients $a/4$, $b/2$ and $c$
are depicted in Fig. 2a, 2b, 2c, respectively.
The same equation holds, with all helicities reversed,
 for $\dot{u}^-_{n,j}$.
 The numerical values of $a$, $b$ and $c$ are the same of the original
helical shell.

To make contact with the issue of intermittency-dissipation
interrelation,
 we shall 
replace the viscous coefficients $D^{\pm}_n$ of equations (\ref{hel},\ref{TREE})
with an ``effective viscosity'' term, ${\cal D}^{\pm}_n$,  which now acquires both
non-trivial dependencies from time and shell indexes. 
It reads for the two cases:
\be
{\cal D}^{\pm}_{n}(t) \equiv  \nu_{\tiny S}(\delta_{n,N}+\delta_{n,N-1})  {{\left| u_n^{\pm}\right|} \over {k_n}}{k_n^2};\;\;
{\cal D}^{\pm}_{n,j}(t) \equiv  \nu_{\tiny S}(\delta_{n,N}+\delta_{n,N-1}) 
 {{\left| u_{n,j}^{\pm}\right|} \over {k_n}}{k_n^2}
\ee
where $\nu_{\tiny S}$ is an empirical constant of order $1$.
This ``sub-grid-scale'' term is clearly patterned after the simplest 
NS effective viscosity model.
The only difference is that due to the
short range interactions of our shell models, the sub-grid-modeling
is applied only to the last and last-but-one shells 
$k_N$, $k_{N-1}$.\\
Our sub-grid closure combines features of 
the classical Smagorinski Large Eddy Simulation
model and the so-called hyperviscosity models used 
in the direct spectral simulation
of incompressible turbulence.
This is consistent with the double-locality in real and momentum space
of
the wavelet basis functions.

The two methods are quite different in scope 
and formulation: Smagorinski works in
real space as a local, dynamic, effective viscosity responding to the
local
stress so as to mimic the effects of unresolved scales on the resolved
ones. 
Hyperviscosity is local in $k$-space, static, and does {\it not} aim at
representing the effects of unresolved scales, but simply at reducing
the
size of the dissipative region so as 
to take full advantage of the grid resolution.

\section{Results}
As previously observed, the common aim of any turbulence model or 
large-eddy simulation is to capture the effects of unresolved scales 
on the resolved ones. In practice, this means that
once the subgrid model is appropriately tuned 
the resolved scales should be basically unaffected by grid resolution
\cite{BORIS}.

This is indeed the case for our sub-grid model.
In Figure 3 we show the energy spectra for the chain model with 
eddy-viscosity at three different resolutions $N=16,20,24$. 
For the sake of comparison the case 
with normal viscosity is also reported for $N=16$. 
As a first remark, we note that the presence of the eddy-viscosity
considerably
widens the inertial regime 
which extends deep down to the last-but one shell.     
Moreover, the slope of the spectrum is basically the same
independent of the number of shells used, which is exactly 
the property we were looking for.

We note that is not the case with normal viscosity, where 
in order to widen the inertial range it is necessary to lower the
value of the viscosity so as to increase the Reynolds number. 
Of course resolution must be increased accordingly so as to resolve the 
dissipative region in order to prevent numerical problems. 
In order to gain a more quantitative assessment on the grid-independence
of our results, we shall evaluate the scaling exponents $\zeta_p$ up to
$p=8$. 
In Table 1 we show the scaling exponents for the chain model 
with ($\zeta_p^S$) and without ($\zeta_p^D$)
sub-grid eddy-viscosity (``S'' stands for sub-grid and ``D'' for
direct). 
The simulation was run with $16$ shells for about $10^5$ eddy turn over
time of the largest scale.\\
The first remark is that in both cases a significant departure
from Kolmogorov K41 law is observed, i.e. the 
sub-grid model does {\bf not} destroy intermittency.\\
More precisely, $\zeta_p^S$ and $\zeta_p^D$ coincide 
within statistical error, which means that  intermittency 
survives and it is basically insensitive to eddy viscosity.
The scaling exponents reported in Table 1 have been computed as a direct
fit on structure functions in $k$ space. Statistical accuracy is
generally
good due to the  large number of shells available.\\
It is nonetheless interesting to note that such an estimate is even more
accurate using Extended Self Similarity (ESS), namely by representing 
the $p-th$ order structure function versus the third order one.
In Figure 4 we show $S_6$ as a function of $S_3$ for the case {\it with}
and {\it without} eddy viscosity. As we see, the case of normal viscosity
displays two distinct slopes in the inertial and dissipative regimes,
whereas
with eddy viscosity this slope is everywhere  the 'inertial' one.\\
This suggests that the combined use of LES models and ESS analysis 
might prove useful for the analysis of scaling exponents 
in more complex simulations. 

We now move on to the discussion of the results with the tree model.

Before analyzing the results it is worth to point out that
the tree formulation makes more contact with the usual Navier-Stokes
 Smagorinski 
eddy viscosity in that
it introduces a spatial dependence in the model.
It is therefore of interest to investigate how this spatial dependence 
is going to affect the physical picture described so far.

The physical picture as it comes from the analysis of
intermittency in the inertial range is pretty much the same as with the
chain model: in particular, intermittency survives 
and shows no dependency on whether a sub-grid closure is used or not
(see Table 2).\\
The actual numerical values of the scaling
exponents are slightly higher than in the chain 
case, and this is hardly surprising since the tree model allows 
for spatial redistribution of the energy flow so that spotty events 
are somehow smeared out.

\subsection{Refined Kolmogorov Hypothesis (RKH)}
In a tree structure we may also test the robustness of  the RKH.
 As previously discussed,
the RKH links statistical properties
of the energy dissipations, $\epsilon(r)$ averaged
on a box of size $r$, to the inertial range fluctuations, $\delta v(r)$:
\be
\epsilon_r(x) = {1 \over {r^3}} \int_{\Lambda_r(x)} 
\epsilon(y) \; d^3y \sim {{(\delta v(r))^3} \over r}.
\end{equation}
In particular one may therefore write:
\begin{equation}
\left\langle \epsilon_r(x)^{p/3} \right\rangle \sim S_{p}(r)/r^{p/3}.
\label{lu}
\end{equation}

The first step in constructing the energy dissipation field
in any tree model \cite{bbtt} is to consider 
the energy dissipation {\it density}, $\eta_{n,j}$,
in the structure  covering the region $\Lambda_j(n)$
of length $2^{-n}$, centered in the spatial
site labeled by $j$.
These structures are represented by boxes in Fig.1.

In the case with eddy viscosity we have
\begin{equation}
\eta_{n,j} = 
{\cal D}^{\pm}_{n,j}\left( \left|u^+_{n,j} \right|^2 + 
\left|u^-_{n,j}\right|^2\right).
\label{eq:eta}
\end{equation}
Let us notice that in the above expression only the 
last and the last-but-one shells give non-zero contribution; at difference
with the case when a molecular viscosity acting on all scales is
considered (the latter would correspond to the choice of $D$ instead of
 ${\cal D}$ in eq. \ref{eq:eta}).

The total energy dissipation density,
$\epsilon = (1/\Lambda_T) \int_{\Lambda_T} \epsilon(x) \, dx $,
where $\Lambda_T$ is the total space length, 
is, by definition,  the sum of all these contributes
(sum over boxes at all scales in Fig. 1):

\begin{equation}
\epsilon = \sum_{n,j} 2^{-n} \eta_{n,j}.
\label{eq:e1}
\end{equation}

On the other hand, in order to study the scaling properties 
of the energy dissipation field, one has to disentangle in 
 $\epsilon$ the  contributions coming 
from the coarse-grained energy dissipation field $\epsilon_r$.

In our formulation, we can then rewrite:

\begin{equation}
\epsilon = \frac{1}{\Lambda_T} \int_{\Lambda_T} \epsilon(x) \, dx =
\frac{1}{2^{n-1}} \sum_{j=1}^{2^{n-1}} 
\left( {\frac{1}{2^{-n}} \int_{\Lambda_j(n)} \epsilon(x) \, dx }\right)=
\frac{1}{2^{n-1}} \sum_{j=1}^{2^{n-1}}  \epsilon_{n,j},
\label{eq:e2}
\end{equation}

where the last expression
 is independent of $n$ and 
 the $\epsilon_{n,j}$'s are the coarse-grained energy
 dissipation densities,
obtained as averages over spatial regions of length $2^{-n}$.
Note that the average density $\epsilon_{n,j}$ over $\Lambda_j(n)$
does not coincide simply with the density $\eta_{n,j}$ of the
structure living in $\Lambda_j(n)$, namely:

\begin{equation}
\epsilon_{n,j} = \eta_{n,j} + \sum_{m < n} \eta_{m,k(m)} +
\sum_{m > n} \left\langle\eta_{m,k(m)}\right\rangle_{I(m)}.
\label{eq:e3}
\end{equation}

Here, in the second (third) term of the RHS we take into account
density contributions coming from larger (smaller) scale structures
(as an example, all regions contributing to the definition
of $\epsilon_{n,j}$ are represented as shadowed boxes in Fig. 1).
The index $k(m)$ in the second term of RHS
 labels the  location of larger scale
structures containing the region  $\Lambda_j(n)$ 
under consideration (shadowed boxes with $m < n$ in Fig. 1). 
In the third term, 
an average  is performed over  
$k(m) \in I(m)$, where $I(m)$ labels the set of structures 
contained in  $\Lambda_j(n)$, for any $m > n$
(in Fig. 1, $I(m)$ labels the two boxes at $n+1$,
the four boxes at $n+2$, and so on).

The best spatially resolved energy dissipation field is for $n=N$:

\begin{equation}
\epsilon_{N,j} = \sum_{m \le N} \eta_{m,k(m)};
 \,\,\,\,\,\,\,\,\,\,\,\,\,\,\,\,\,\,\,\,\,\,\,\,\,\,\, j=1,...,2^{N-1}.
\label{eq:e(x)}
\end{equation}

In Fig. 5,  the instantaneous values assumed by
$\epsilon_{N,j}$ in the $N_T/2=32768$ locations of the last 
level are showed.
The chaotic, intermittent character of this spatial
signal is evident.

In Table 3 we show that the RKH is 
still well verified also in the sub-grid modeling picture, proving
to be a robust and non-trivial property connecting small
scales and inertial range scales in turbulent flows.  


\section{Conclusions}

Summarizing, we have presented a detailed
study of dynamical eddy-viscosity models in chain and tree shell models
of fluid turbulence. \\
The main goal was to check whether or not the inertial range
properties are affected by the way the flow dissipate
energy. We found a strong robustness of inertial range intermittency
once the proposed eddy-viscosity model is implemented
in our shell models. \\ The eddy-viscosity closure that we 
have adopted may also be regarded as a multiplicative closure
of the small-scales equations of motion, i.e. it is tantamount to assuming
that $ u_{n+1} \sim a_{n+1,n} \cdot u_{n}$ with an appropriate 
multiplicative random coefficient $a_{n+1,n}$. 
The fact that intermittency is not affected by the details
of the eddy-viscosity models indicates that fine-tuning of the 
coefficients in front to the eddy-viscosity term is probably not demanded. 
Nevertheless, oversimplified eddy-viscosity models
based only on dimensional analysis
would probably fail on the same goal, due to their inability to 
dissipate violent intermittent bursts.\\
Moreover, the usual phenomenological RKH which links
energy dissipation statistics with inertial range properties
is also largely unaffected by this kind of modeling.\\ 
Whether the same universality is present in real Navier-Stokes
equations is still a matter of debate in the scientific
community  \cite{chen,orszag2d,eyink}. Certainly, in order to properly
test this question it is always necessary a fine resolution
of the smallest resolved scales and, more important, a detailed
study of the dependence on finite Reynolds effects. Indeed,
in many cases, bottlenecks phenomena close to the sub-grid modeling
scales may arise \cite{lohse}. 
These bottleneck effects may introduce a finite-Reynolds 
bias which could lead to erroneous conclusions on the dependence of inertial
range statistics on eddy viscosity or hyperviscosity modeling.

We acknowledge the help of E. Trovatore for making some of the figures
and for sharing with us  her  numerical results. \\
L.B and F.T have been partially supported by INFM (PRA-TURBO).

\newpage
\section*{Table Captions}
\begin{enumerate}
\item [Table 1]  Scaling exponents  for the chain 
model with eddy viscosity, $\zeta^S(p)$,  
for $N=16,20,24$ and with normal viscosity, $\zeta^D(p)$,  with $N=16$. 

\item [Table 2] Scaling exponents for the tree model {\it with} eddy
viscosity $\zeta^S(p)$,  and
{\it without} eddy viscosity, $\zeta^D(p)$.

\item [Table 3] Slope, $\chi(p)$, of the log-log plot of 
 equation (\ref{lu}) for the tree model for $p=1, \dots, 10$. Notice
that when $\chi(p)=1$ the RKH is verified.
\end{enumerate}
\section*{Figure Captions}

\begin{enumerate}

\item [Fig. 1] A picture of the hierarchical system, covering the 
                one-dimen\-sio\-nal interval $[0,\Lambda_T]$.

\item [Fig. 2] Type of interaction (a, b and c) for the tree model.

\item [Fig. 3] Log-log plot of the  energy  spectra versus the wavenumber 
for the chain 
model with eddy-viscosity at three different resolutions $N=16$ (pluses),
$N=20$ (stars), $N=24$ (crosses). 
For the sake of comparison the case 
with normal viscosity is also reported for $N=16$ (dotted line).
The straight line has slope $-1-\zeta_2$.

\item [Fig. 4]  Log-log plot of  $S_6$ 
versus $S_3$ for the chain model with $N=16$ {\it with}
eddy viscosity  (pluses) and {\it without} eddy-viscosity (crosses).
The straight line has slope $\zeta_6$.

\item [Fig. 5] Instantaneous configuration of the coarse-grained 
                energy dissipation
                density field, $\epsilon_{N,j}$, over the last level sites.
\end{enumerate}
\clearpage
%
%
\begin{table}
\begin{center}
\begin{tabular}{|r|r|r|r|r|}

\hline
p & $\zeta^D(p)$ & $\zeta^S(p)$ & $\zeta^S(p)$ & $\zeta^S(p)$ \\

  & $N=16$  & $N=16$   & $N=20$   & $N=24 $  \\
\hline
 1 & $0.368 \pm 0.007$ & $0.367 \pm 0.002$& $0.367 \pm 0.002$&$0.367 \pm 0.001$ \\ \hline
 2 & $0.700 \pm 0.005$ & $0.699 \pm 0.002$& $0.699 \pm 0.002$&$0.699 \pm 0.001$ \\ \hline
 3 & $1.0 \pm 0.0$ & $1.0 \pm 0.0$& $1.0 \pm 0.0$&$1.0 \pm 0.0$ \\ \hline
 4 & $1.271 \pm 0.007$ & $1.273 \pm 0.004$& $1.268 \pm 0.007$&$1.272 \pm 0.003$ \\ \hline
 5 & $1.52 \pm 0.01$ & $1.522 \pm 0.007$& $1.50 \pm 0.02$&$1.518 \pm 0.008$ \\ \hline
 6 & $1.74 \pm 0.02$ & $1.75 \pm 0.01$& $1.71 \pm 0.04$&$1.74 \pm 0.02$ \\ \hline
 7 & $1.94 \pm 0.04$ & $1.97 \pm 0.01$& $1.90 \pm 0.07$&$1.96 \pm 0.02$ \\ \hline
 8 & $2.12 \pm 0.05$ & $2.17 \pm 0.02$& $2.08 \pm 0.09$&$2.16 \pm 0.03$ \\ \hline
 9 & $2.29 \pm 0.08$ & $2.37 \pm 0.02$& $2.3 \pm 0.1$&$2.36 \pm 0.04$ \\ \hline
10 & $2.5 \pm 0.1$ & $2.57 \pm 0.03$& $2.4 \pm 0.1$&$2.56 \pm 0.04$ \\ \hline
11 & $2.6 \pm 0.1$ & $2.76 \pm 0.04$& $2.6 \pm 0.1$&$2.76 \pm 0.05$ \\ \hline
12 & $2.8 \pm 0.2$ & $2.96  \pm  0.06$& $2.8 \pm 0.2$&$2.96 \pm 0.06$ \\ \hline
\end{tabular}
\label{tab:tree1}
\caption{}

\end{center}
\end{table}

\begin{table}
\begin{center}
\begin{tabular}{|r|r|r|}
\hline
p &$ \zeta^D(p)$  &  $\zeta^S(p)$ \\
 
\hline
 1 & $0.348 \pm 0.005$ & $ 0.347 \pm 0.005$ \\ \hline
 2 & $0.682 \pm 0.005$ & $ 0.681 \pm 0.005$  \\ \hline
 3 & $1.00$            & $ 1.00 $  \\ \hline
 4 & $1.303 \pm 0.006$ & $ 1.302 \pm 0.006$  \\ \hline
 5 & $1.59 \pm 0.01$   & $ 1.59  \pm 0.01$  \\ \hline
 6 & $1.86 \pm 0.02$   & $ 1.85  \pm 0.02$  \\ \hline
 7 & $2.12 \pm 0.03$   & $ 2.10  \pm 0.03$  \\ \hline
 8 & $2.35 \pm 0.03$   & $ 2.32  \pm 0.03$  \\ 
\hline 
\end{tabular}
\label{tab:tree}
\caption{}
\end{center}
\end{table}

\begin{table}
\begin{center}
\begin{tabular}{|r|r|}
\hline
p &  $\chi(p)$    \\
 
\hline
 1 & $1.00 \pm 0.02$  \\
\hline
 2 & $1.001 \pm 0.008$  \\
\hline
 4 & $1.000 \pm 0.007$  \\
\hline
 5 & $1.000 \pm 0.01$  \\
\hline
 6 & $1.00 \pm 0.02$  \\
\hline
 7 & $1.01 \pm 0.03$  \\ 
\hline
 8 & $1.02 \pm 0.04$  \\
\hline
 9 & $1.02 \pm 0.06$  \\
\hline
 10 & $1.02 \pm 0.07$ \\
\hline 
\end{tabular}
\label{tab:rksh}
\caption{}
\end{center}
\end{table}

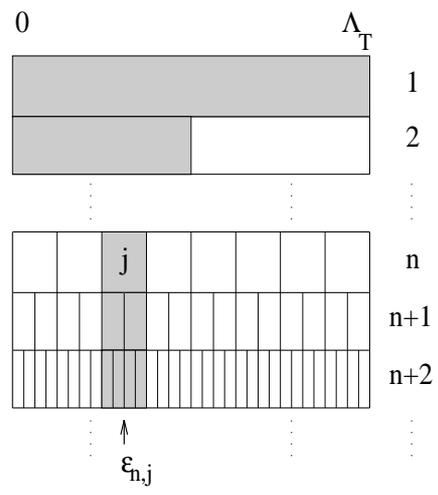
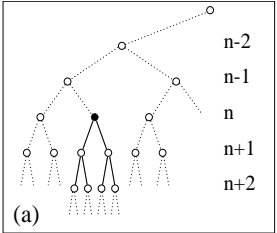
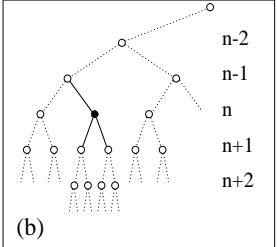
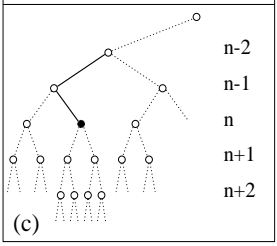
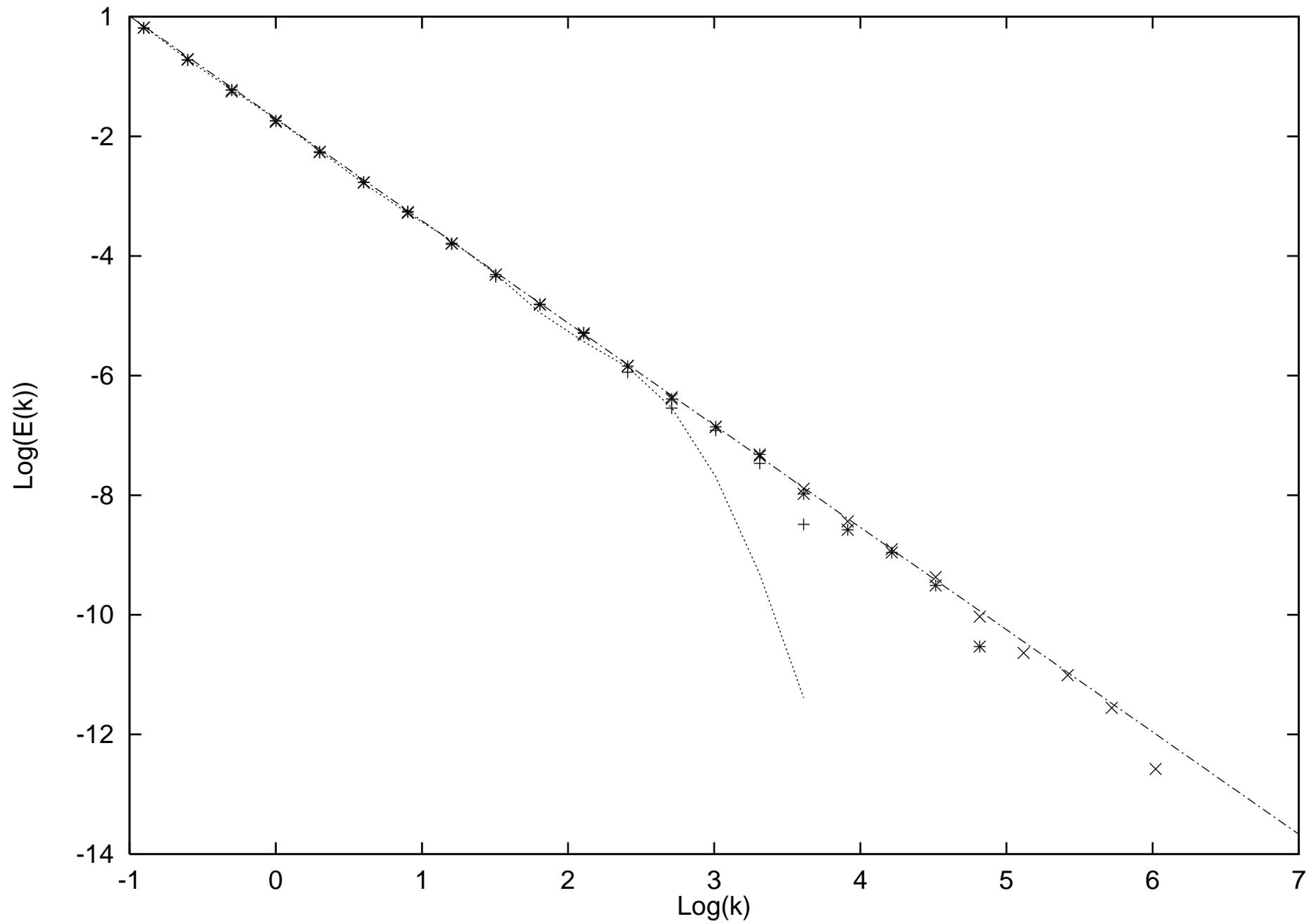
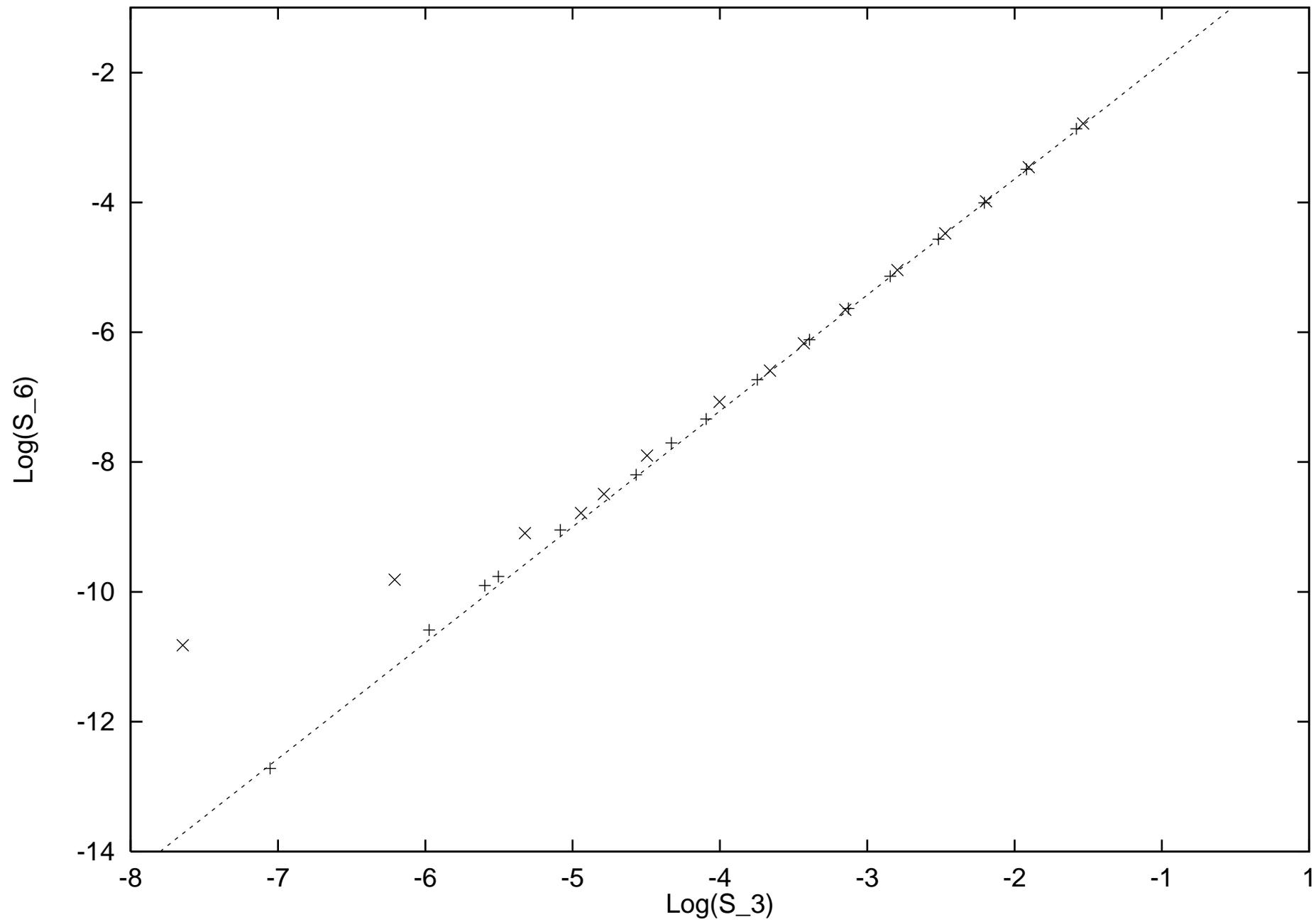
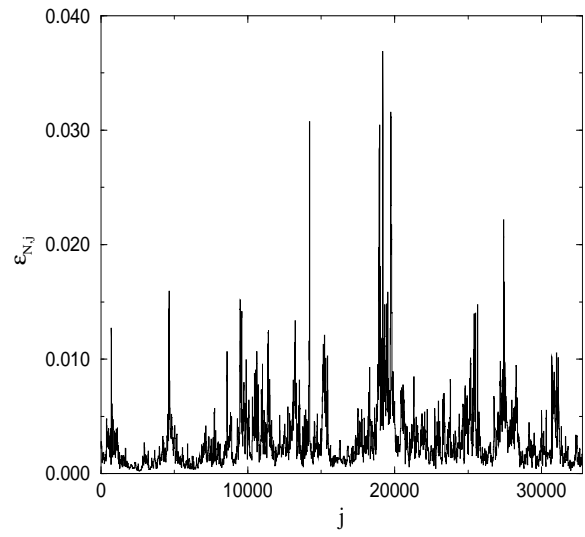
\end{document}